\documentclass[final,3p,times,twocolumn,authoryear]{elsarticle}

\usepackage{float}
\usepackage{pifont}
\usepackage{booktabs}
\usepackage{glossaries}

\glsdisablehyper
\usepackage{csquotes}
\makeglossaries
\usepackage{subcaption}
\usepackage{comment}
\usepackage[most]{tcolorbox}
\usepackage{xcolor}
\usepackage{soul}
\usepackage{hyperref}
\sethlcolor{yellow} 
\usepackage{lipsum}

\usepackage{tabularx}
\usepackage{booktabs}
\usepackage{natbib}
\setcitestyle{square}
\setcitestyle{numbers}

\begin{document}

\begin{frontmatter}
\title{SynthCTI: LLM-Driven Synthetic CTI Generation to enhance MITRE Technique Mapping}

\author[label1]{Álvaro Ruiz-Ródenas\corref{cor1}}
\ead{a.ruizrodenas@um.es}

\author[label1]{Jaime Pujante Sáez}
\ead{jaime.pujantes@um.es}

\author[label2]{Daniel García-Algora}
\ead{d.galgora@upm.es}

\author[label1]{Mario Rodríguez Béjar}
\ead{mario.rodriguezb1@um.es}

\author[label2]{Jorge Blasco}
\ead{jorge.blasco.alis@upm.es}

\author[label1]{José Luis Hernández-Ramos}
\ead{jluis.hernandez@um.es}

\affiliation[label1]{organization={Department of Information and Communication Engineering, Universidad de Murcia},
            postcode={30100},
            city={Murcia},
            country={Spain}}
\affiliation[label2]{organization={Department of Computer Systems, Universidad Politécnica de Madrid},
            postcode={28031},
            city={Madrid},
            country={Spain}}

\cortext[cor1]{Corresponding author}

\begin{abstract}
Cyber Threat Intelligence (CTI) mining involves extracting structured insights from unstructured threat data, enabling organizations to understand and respond to evolving adversarial behavior. A key task in CTI mining is mapping threat descriptions to MITRE ATT\&CK techniques. However, this process is often performed manually, requiring expert knowledge and substantial effort. Automated approaches face two major challenges: the scarcity of high-quality labeled CTI data and class imbalance, where many techniques have very few examples. While domain-specific Large Language Models (LLMs) such as SecureBERT have shown improved performance, most recent work focuses on model architecture rather than addressing the data limitations. In this work, we present SynthCTI, a data augmentation framework designed to generate high-quality synthetic CTI sentences for underrepresented MITRE ATT\&CK techniques. Our method uses a clustering-based strategy to extract semantic context from training data and guide an LLM in producing synthetic CTI sentences that are lexically diverse and semantically faithful. We evaluate SynthCTI on two publicly available CTI datasets, CTI-to-MITRE and TRAM, using LLMs with different capacity. Incorporating synthetic data leads to consistent macro-F1 improvements: for example, ALBERT improves from 0.35 to 0.52 (a relative gain of 48.6\%), and SecureBERT reaches 0.6558 (up from 0.4412). Notably, smaller models augmented with SynthCTI outperform larger models trained without augmentation, demonstrating the value of data generation methods for building efficient and effective CTI classification systems.

\end{abstract}


\end{frontmatter}

\section{Introduction}

Cyber Threat Intelligence (CTI) has become an essential component of proactive cybersecurity strategies by providing actionable insights about ongoing and emerging threats \cite{alaeifar2024current}\cite{brown2023sans}. Despite its growing relevance, \textit{CTI mining} remains a challenging task due to the unstructured nature of threat data, domain-specific vocabulary, and the evolving tactics of adversaries \cite{sun2023cyber}. To structure and translate these insights into actionable knowledge,  the MITRE ATT\&CK framework\footnote{https://attack.mitre.org/} provides a structured taxonomy of adversarial tactics and techniques based on real-world observations, supporting the alignment of CTI content with known threat behaviors. However, the task of mapping unstructured CTI reports to specific ATT\&CK techniques remains largely manual, requiring expert domain knowledge and significant time investment \cite{della2025cti}.

Existing methods for automating this mapping (ranging from traditional \gls{ml} and \gls{dl} models, and more recently, \gls{llm}), face two main key limitations. First, there is a lack of high-quality labeled \gls{cti} data, which restricts the training of robust classifiers. Second, available datasets (such as CTI-to-MITRE \cite{ctidataset} and \gls{tram} \cite{tramdataset}) suffer from extreme class imbalance, where certain techniques are well-represented and others have very few samples. While recent domain-specific models like SecureBERT \cite{aghaei2023securebert} have shown performance improvements in such datasets compared to more traditional \gls{ml} approaches \cite{orbinato2022automatic}, most efforts have focused on model architecture rather than addressing the data scarcity challenge \cite{li2024automated}\cite{you2024cyber}. Moreover, conventional data augmentation techniques often fail to capture the semantic specificity required in the cybersecurity domain, limiting their effectiveness. In this context, synthetic data generation via \glspl{llm} emerges as a promising solution \cite{li2024empowering}. \glspl{llm} can learn domain-specific linguistic patterns and produce contextually rich, coherent text. Indeed, recent approaches offer potential for synthetic data generation, but their application in CTI remains underexplored and often lacks domain-specific tailoring \cite{li2024empowering}\cite{dai2025auggpt}.

In this work, we address these challenges by introducing \textit{SynthCTI}, a synthetic data augmentation framework that leverages \glspl{llm} to enrich training data for underrepresented MITRE techniques. While most previous work does not incorporate data augmentation, a few exceptions apply it indirectly through a prompt-based approach \cite{bayer2022multi}. In contrast, SynthCTI builds detailed, semantically guided prompts using contextual features extracted from clustered CTI examples. This targeted approach enables the generation of high-quality, diverse sentences that align with cybersecurity semantics and preserve class-specific nuances. Our approach begins by clustering sentence embeddings using HDBSCAN \cite{Campello2015} to identify semantically coherent subgroups within each MITRE technique. From these clusters, we extract contextual features, such as LDA-derived topics, KeyBERT keywords, tone, and representative examples, to construct informative prompts. These prompts are then used to guide a LLM, in synthesizing CTI sentences that are both semantically faithful and linguistically diverse. The resulting synthetic examples are incorporated into the training data to improve classifier performance.

We evaluate SynthCTI on two publicly available datasets (CTI-to-MITRE and \gls{tram}) demonstrating its effectiveness across models of diverse capacity. Incorporating synthetic data leads to substantial macro-F1 gains. For instance, SecureBERT improves from 0.4412 to 0.6558, while even lightweight models such as DistilBERT and Albert \cite{Albert} achieve competitive performance that improves larger models trained without augmentation. This is particularly relevant for real-world deployment scenarios where smaller models provide advantages in terms of computational efficiency, inference speed, and privacy preservation. These results highlight the effectiveness of SynthCTI and its potential to enable more efficient and deployable solutions in \gls{cti} classification.

In summary, the main contributions are:
\begin{itemize}
    \item We build a synthetic data generation pipeline tailored to enhance the classification of \gls{cti} into MITRE ATT\&CK techniques.
    
    \item We develop a prompt-engineering strategy combining HDBSCAN clustering, topic modeling, keyword extraction, and tone analysis to guide LLM-based generation.
    
    \item We show that guided generation produces high-quality, semantically coherent \gls{cti} examples that enrich underrepresented classes.

    \item We perform an empirical validation across two real-world \gls{cti} datasets, demonstrating substantial macro-F1 gains.
\end{itemize}

The rest of the paper is structured as follows: Section~\ref{sec:related} reviews related work on \gls{cti} classification and LLM-based data augmentation. Section~\ref{sec:design} details our proposed SynthCTI pipeline. Section~\ref{sec:evaluation} presents experimental results and in-depth analysis of augmentation quality and classifier performance. Section~\ref{sec:conclusion} concludes the paper.

\newacronym[plural=IOCs, firstplural=Indicators of Compromise]{ioc}{IOC}{Indicators of Compromise}
\newacronym[plural=APTs, firstplural=Advanced Persistent Threats]{apt}{APT}{Advanced Persistent Threat}
\newacronym{ttp}{TTP}{Tactics, Techniques and Procedures}
\newacronym{ml}{ML}{Machine Learning}
\newacronym{nlp}{NLP}{Natural Language Processing}
\newacronym{dl}{DL}{Deep Learning}
\newacronym{bilstm}{BiLSTM}{Bidirectional Long Short-Term Memory }
\newacronym{cti}{CTI}{Cyber Threat Intelligence}
\newacronym{ctr}{CTR}{Cyber Threat Report}
\newacronym[plural=LLMs, firstplural=Large Language Models]{llm}{LLM}{Large Language Model}
\newacronym{lstm}{LSTM}{Long Short-Term Memory}
\newacronym{elmo}{ELMo}{Embeddings from a Language Model}
\newacronym{svm}{SVM}{Support Vector Machines}
\newacronym{mlp}{MLP}{Multi-Layer Perceptron}
\newacronym[plural=CCNs, firstplural=Convolutional Neural Networks]{cnn}{CNN}{Convolutional Neural Network}
\newacronym{ner}{NER}{Named-Entity Recognition}
\newacronym{hdbscan}{HDBSCAN}{Hierarchical Density-Based Spatial Clustering of Applications with Noise}
\newacronym{dbscan}{DBSCAN}{Density-Based Spatial Clustering of Applications with Noise}
\newacronym{ctid}{CTID}{Center for Threat-Informed Defense}
\newacronym[plural=LLMs, firstplural=Large Language Models]{llm}{LLM}{Large Language Models}
\newacronym{lda}{LDA}{Latent Dirichlet Allocation}
\newacronym{tram}{TRAM}{Threat Report ATT\&CK Mapper}

\section{Related work}\label{sec:related}
Recent work has explored the use of \gls{ml} and \glspl{llm} to improve the automated classification of \gls{cti} text into MITRE ATT\&CK tactics and techniques. Earlier approaches focused on traditional \gls{nlp} and \gls{ml} techniques, including TF-IDF representations combined with classifiers such as \gls{svm}, decision trees and logistic regression \cite{ayoade2018automated} \cite{legoy2020automated}. The authors of \cite{orbinato2022automatic} performed a comparison of traditional and \gls{dl} classifiers for this purpose. They experimented with several types of models, including logistic regression, \glspl{svm}, \gls{lstm}, \glspl{cnn}, and SecureBERT~\cite{aghaei2023securebert}, a BERT model pre-trained on cybersecurity text. SecureBERT outperformed all baselines on a dataset of 12,945 CTI samples, labeled with 188 MITRE techniques, named CTI-to-MITRE. Expanding this line of work, \cite{li2024automated} proposed a sentence-level classification pipeline using DistilBERT for multi-label classification of both tactics and techniques. Their framework includes a post-processing step that corrects label assignments based on the structural hierarchy of the ATT\&CK framework. Evaluated on over 26,000 CTI sentences, their proposed method outperformed traditional methods.

\cite{you2024cyber} introduced CTI-BERT, a BERT model fine-tuned on domain-specific data using a two-stage training process. They first trained on ATT\&CK descriptions, then fine-tuned on a labeled subset of the \gls{tram} dataset. This study shows the importance of domain adaptation and fine-tuning when applying LLMs to complex tasks such as CTI classification. Furthermore, the authors of \cite{zhang-etal-2025-evaluating} also show this need by benchmarking LLMs on several domain-specific datasets, including CTI-to-MITRE and \gls{tram} datasets, showing significant variance in performance across models and the importance of task-specific evaluation frameworks for enterprise applications.

The development and availability of pre-trained generative models, such as BERT, GPT and T5, have significantly enhanced text augmentation capabilities through synthetic generation. These models are able to generate complete sentences or even entire documents that maintain coherence and contextual relevance \cite{feng2021}. Although these methods have distinct advantages in terms of semantic richness and diversity generated, they need careful validation and quality control procedures to ensure that no distortions are introduced that could negatively affect the performance of the final models.

The authors of ~\cite{cuong2025towards} took a hybrid approach, using GPT-3.5-based summarization with a SciBERT ~\cite{beltagy2019scibert} (a BERT-like model focused on scientific text) classifier. In this pipeline, long \gls{cti} documents are first summarized using a generative LLM, and then classified into ATT\&CK techniques. They also used GPT-3.5 to augment their training data by generating synthetic \gls{cti} sentences for underrepresented techniques, although, the authors fail to explain their process to generate new data. Similarly, in ~\cite{bayer2022multi}, the authors make use of LLM-generated data augmentation through a pipeline combining GPT-3 text synthesis, human-guided filtering, and few-shot learning. They applied their pipeline to cybersecurity-related tweets, to determine if they are \gls{cti} related or not, improving F1 scores by 21 points over standard training, and by 18 points over baseline few-shot methods. However in their ablation study, they demonstrated that most of the improvement comes from their multi level fine-tunning and the impact of their method for data augmentation is minimal.

Our work is focused on using synthetic data to improve the automated classification of \gls{cti} sentences into  MITRE ATT\&CK techniques. We compare several augmentation techniques and propose a data augmentation pipeline  that involves clustering, contextual feature extraction, and prompt-based generation. This process allows us to capture the semantic structure of underrepresented classes and use that information to guide the generation of synthetic data.. For our method, we use Gemma 3 \cite{gemma3} to generate new data based on information, such as key words and tone, extracted from the least represented classes in the dataset and examples from those classes, unlike \cite{bayer2022multi} where the authors only use a few-shot approach giving the model a few examples of each class to generate the new data. We leverage this feature extraction to generate context for the generation of new data, with the goal of obtaining new high quality data.  To validate the proposed method, several classification models based on \glspl{llm} have been used, including ALBERT, DistilBERT, BERT and SecureBERT.

\section{System overview}\label{sec:design}

\begin{figure*}[h]
    \centering
    \includegraphics[trim=0 60 0 10, clip, width=1\linewidth]{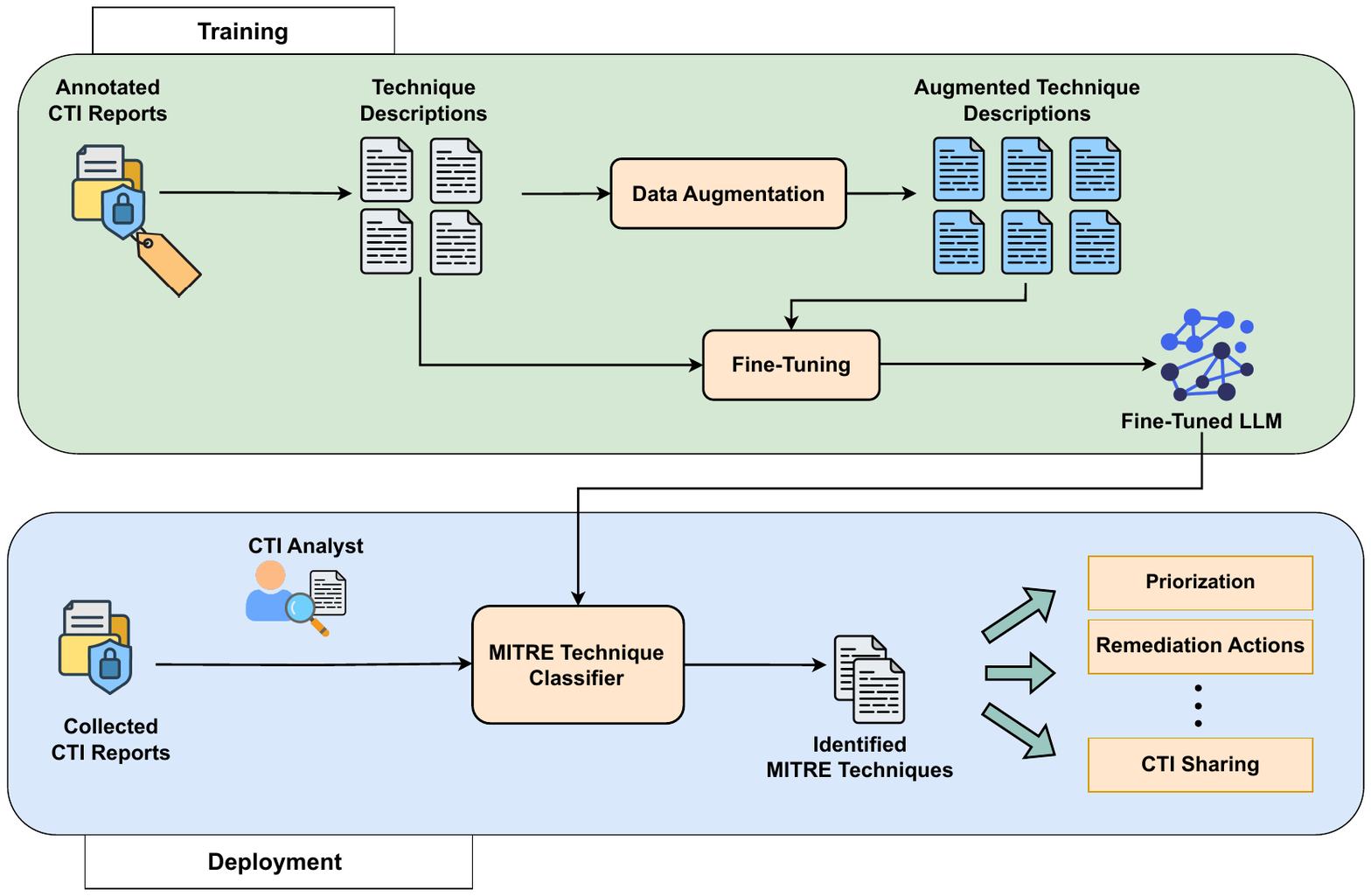} 
    \caption{General overview of \textit{SynthCTI}.} 
    \label{fig:Overview} 
\end{figure*}

Figure~\ref{fig:Overview} shows an overview of our \textit{SynthCTI} framework, which is divided into two main phases. During \emph{System Training}, we address the data imbalance in \gls{cti} sentence classification by increasing underrepresented classes using \glspl{llm}. In particular, the training data is partitioned by class. Then, each class subset is transformed into embeddings and clustered to identify semantically coherent subgroups. For each cluster, key features are extracted, including representative sentences, central topics, and phrases enriched with contextually appropriate synonyms. These features are used to construct structured prompts that guide a \gls{llm} in generating synthetic sentences that align with the semantic and stylistic characteristics of the original data. The resulting synthetic samples are then combined with the original training data to fine-tune a collection of pre-trained models for multi-class classification.

In the \emph{System Deployment} phase, the fine-tuned models assist \gls{cti} analysts in mapping sentences from new \gls{cti} reports to their corresponding MITRE ATT\&CK techniques. Once techniques are identified, analysts, or automated workflows integrated with the system, can generate specific recommendations, prioritize threats, and formulate targeted mitigation strategies.

\begin{figure*}[h]
    \centering
    \includegraphics[trim=20 280 40 120, clip, width=1\linewidth]{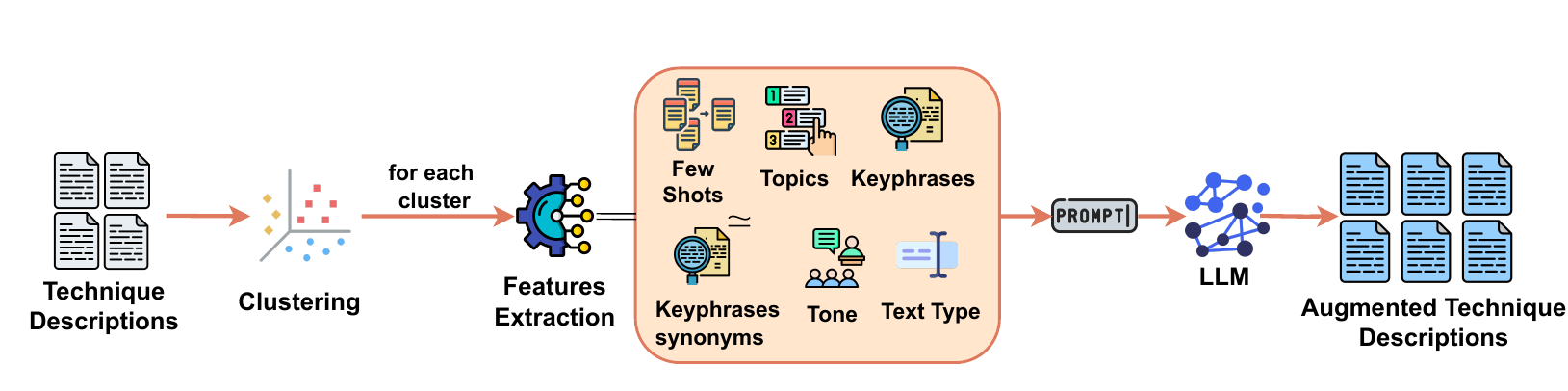} 
    \caption{General view of the Data Augmentation process.} 
    \label{fig:data_aug} 
\end{figure*}

\subsection{System Training}

The training process begins with a collection of annotated \gls{cti} sentences, labeled with a specific MITRE ATT\&CK technique. These annotations are typically derived from unstructured intelligence sources, such as threat reports or incident analyses, where adversary behaviors are described in natural language \cite{you2024cyber}. A common issue of \gls{cti} corpora is their highly skewed distribution. While some techniques are represented by hundreds of samples, others have only a few due to limited reporting or their niche application. This imbalance negatively impacts the classifier's ability to generalize across all techniques. To mitigate this, we propose a data augmentation strategy tailored to enrich sparsely represented techniques with realistic synthetic samples. This strategy is further described below. 

\subsubsection{Data augmentation}\label{dataug}
We implement a data augmentation strategy to generate synthetic \gls{cti} sentences using \glspl{llm}. Rather than producing generic text, our approach builds prompts grounded in the internal semantic structure of each technique class. This process is described in more detail in Figure~\ref{fig:data_aug}. To achieve this, we first cluster training samples within each class using sentence embeddings. Then, for each cluster, we extract key semantic and linguistic features to construct prompts tailored to its content. These prompts guide the LLM in generating synthetic sentences that preserve the topical and linguistic characteristics of the original data.

The process consists of three main steps:
1) semantic clustering of sentence embeddings; 2) extraction of representative features, and 3) construction of \gls{llm} prompts based on those features. We describe each step in detail below.

\paragraph{Clustering}
We begin by transforming the \gls{cti} sentences into vector representations using a pretrained sentence embedding model. In particular, we use 
the \textit{all-MiniLM-L6-v2}, which represents a lightweight variant of MiniLM~\cite{Wang2020}, fine-tuned following the Sentence-BERT methodology~\cite{Reimers2019}, to allow efficient and semantically meaningful sentence embeddings.

These embeddings are then clustered using \gls{hdbscan} \cite{Campello2015}, a density-based algorithm particularly suited for high-dimensional semantic spaces. Unlike methods such as k-means, \gls{hdbscan} does not require specifying the number of clusters in advance; instead, it identifies groups based on local density. Because it extends \gls{dbscan} \cite{Ester1996}, it can label low-density points as noise, preventing semantically inconsistent samples from contaminating clusters. Additionally, its hierarchical structure allows it to detect clusters with varying densities. This aspect represents a desirable property given the thematic variability often found in \gls{cti} descriptions.

After the application of \gls{hdbscan}, each resulting cluster captures a coherent subset of sentences that reflect specific variations in meaning, tone, or topic focus within a technique class. These clusters form the basis for the next step, where representative features are extracted to build the \gls{llm} prompts.

\paragraph{Feature extraction}
We extract descriptive and structural features from each cluster to construct the prompt that guides the \gls{llm} in generating synthetic \gls{cti} sentences. In particular, we extract the following:

\begin{itemize}
    \item \textbf{Few Shots:} we select a small set of representative examples from each cluster to include in the prompt. Sentences are sorted by their cluster membership probability, computed by \gls{hdbscan}, and the top two are chosen. This few-shot style helps condition the \gls{llm} with context while keeping the prompt size manageable.

   \item \textbf{Topics:}  to extract latent topics, we apply \gls{lda} ~\cite{lda} over a TF-IDF-weighted representation of the cluster. \gls{lda} models each cluster as a mixture of topics, where each topic is a probability distribution over words. We use the top terms from the dominant topics as input hints in the prompt, helping guide the \gls{llm} toward semantically relevant vocabulary and structure.

   \item \textbf{Keyphrases:} to extract keywords, we use KeyBERT~\cite{keybert}, which identifies phrases that are semantically representative of the cluster using contextual embeddings. Unlike frequency-based approaches such as TF-IDF or Bag-of-Words, KeyBERT leverages contextual similarity to rank candidate phrases. The highest-ranked keyphrases are incorporated into the prompt to anchor the \gls{llm} generation on the core concepts of the cluster.

   \item \textbf{Synonyms Keyphrases}: to introduce lexical variety in the generated text, we expand each keyphrase with semantically similar synonyms using WordNet~\cite{Miller1995}, a lexical database that groups words by meaning and grammatical category. For each keyword, we retrieve candidate synonyms and compute a score based on two factors: 1) cosine similarity between the embeddings of the keyword and its synonym, and 2) its usage frequency in natural language, estimated via Zipf scores \cite{moreno2016large}. The top three scoring synonyms are selected to enrich the prompt while keeping it concise. In particular:
   
   For each keyword \textit{w} and each synonym \textit{s}$ \in \mathrm{Syn}(w)\subseteq \mathrm{WordNet}(w)$, 
    
    \[
    \mathrm{score}(w, s) = \underbrace{\cos\bigl(\mathbf{e}_w,\mathbf{e}_s\bigr)}_{\substack{\text{semantic similarity}\\\text{between embeddings}}}
    + \alpha \underbrace{f(s)}_{\substack{\text{usage frequency}\\\text{(Zipf score)}}}
    \]
    
    where $\mathbf{e}_w,\mathbf{e}_s$ are the normalized embeddings of \textit{w} and $\quad
    f(s)$ is the Zipf frequency of \textit{s},
    
    $\alpha$ is a weighting parameter controlling the influence of frequency.

    \item \textbf{Tone:} to estimate the general tone of each cluster, we apply two standard readability metrics that reflect the complexity and formality of the text. 
     
    The Flesch Reading Ease \cite{Flesch} score is inversely proportional to difficulty: lower values indicate dense or technical writing, while higher values suggest simpler, more informal text. Based on this, we define three tone categories: formal (< 30), neutral (30–60), and informal (> 60).

    The second metric is the Gunning Fog Index \cite{Gunning}, which estimates the years of education required to understand the text. Scores above 12 indicate technical content, 9–12 suggest a neutral register, and below 9 imply informal writing.
    
    Each sentence in the cluster is assigned a tone label from both metrics. To derive the cluster’s overall tone, we apply a majority rule: if the most common label exceeds the second most frequent by at least 20\%, it is selected as the representative tone; otherwise, both are retained to reflect linguistic diversity within the cluster.

    \item \textbf{Text Type:} to estimate the structural characteristics of each cluster, we compute the average number of sentences per instance. Longer, multi-sentence texts typically indicate technical descriptions, while shorter ones often reflect a more informal or concise reporting style. This information is included in the prompt to help the LLM reproduce natural variation in length and structure, rather than generating uniformly formatted outputs.

\end{itemize}

\paragraph{LLM Prompt} 
Once the relevant features have been extracted from each cluster, we construct a structured prompt that guides the \gls{llm} in generating new synthetic sentences. The prompt combines representative examples, thematic keywords, and lexical variations to encourage coherent and contextually grounded outputs.

Figure~\ref{fig:data_aug_prompt} illustrates a representative example of the prompt generated for label \textit{T1006} belonging to the \textit{CTI-to-MITRE} dataset\cite{ctidataset}, using the method described above. In particular, the \textit{Examples} section includes representative cluster sentences that mention specific techniques and tools (e.g., "This technique bypasses Windows file access controls..." and "Utilities, such as NinjaCopy..."), offering grounded context to guide generation. The \textit{Key Topics}, extracted via \gls{lda}, highlight relevant concepts such as "file monitoring", "bypass", and "PowerShell", pointing the \gls{llm} to key thematic elements. The \textit{Keyphrases} (e.g., "access controls", "monitoring tools", "NinjaCopy exist") reinforce core lexical items, while the \textit{Synonyms Keyphrases} (e.g., "tool", "utility", "approach") introduce controlled variability in vocabulary. The instruction at the end specifies the generation of 10 sentences with a balanced tone (formal + neutral), aligning the output with the linguistic traits of the cluster. The specific number of sentences generated for a particular technique can be adapted to the required scenario. We provide more details on how we calculate this number for our problem in Section \ref{sec:augmentationcomparison}.

\begin{figure}[ht]
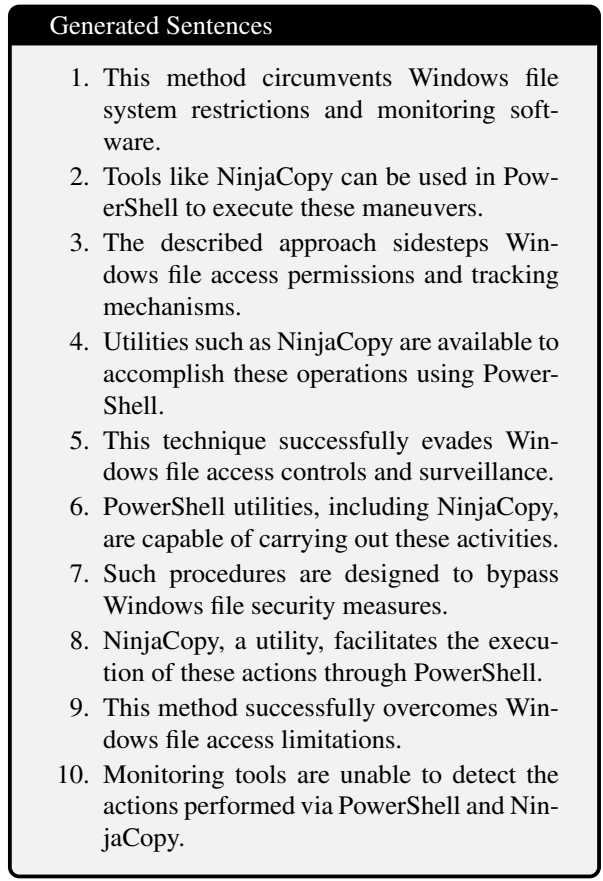

    \centering
    \begin{tcolorbox}[ colback=gray!10, colframe=black, title=Generated Sentences]
    
    \begin{enumerate}
    \item This method circumvents Windows file system restrictions and monitoring software.
    \item Tools like NinjaCopy can be used in PowerShell to execute these maneuvers.
    \item The described approach sidesteps Windows file access permissions and tracking mechanisms.
    \item Utilities such as NinjaCopy are available to accomplish these operations using PowerShell.
    \item This technique successfully evades Windows file access controls and surveillance.
    \item PowerShell utilities, including NinjaCopy, are capable of carrying out these activities.
    \item Such procedures are designed to bypass Windows file security measures.
    \item NinjaCopy, a utility, facilitates the execution of these actions through PowerShell.
    \item This method successfully overcomes Windows file access limitations.
    \item Monitoring tools are unable to detect the actions performed via PowerShell and NinjaCopy.
    \end{enumerate}
    
    \end{tcolorbox}
    \caption{Generated text in response to a prompt about file access techniques.}
    \label{fig:data_aug_output} 
\end{figure}

Using the prompt described above, the Gemma-3 4B \cite{gemma3} model generates the output shown in Figure~\ref{fig:data_aug_output}. The resulting sentences accurately reflect the key concepts provided in the prompt, such as file access controls, monitoring tools, PowerShell, and NinjaCopy, and exhibit lexical variation through the use of synonyms like "utility", "tool" and "approach". While maintaining thematic consistency, the sentences also vary in tone and structure, aligning with the prompt's instruction to mix formal and neutral styles. This example demonstrates the method’s ability to produce diverse yet coherent synthetic samples that preserve the original cluster’s intent.

\begin{figure}[ht]
    \centering
    \begin{tcolorbox}[ colback=gray!10, colframe=black, title=Generated prompt]
    
    \textbf{Examples}  
    
    - This technique bypasses Windows file access controls as well as file system monitoring tools.  
    
    - Utilities, such as NinjaCopy, exist to perform these actions in PowerShell.
    
    \vspace{1em}
    \textbf{Key Topics}  
    
    - \textbf{Topic 0}: file, bypasses, file monitoring, bypasses windows, technique  
    
    - \textbf{Topic 1}: perform actions, actions, exist perform, PowerShell, exist
    
    \vspace{1em}
    \textbf{Keyphrases}  
    
    access controls, actions PowerShell, bypasses windows, controls file, exist perform, file access, file monitoring, monitoring tools, NinjaCopy exist, perform actions, technique bypasses, utilities NinjaCopy, windows file
    
    \vspace{1em}
    \textbf{Synonyms Keyphrases}  
    
    instrument, tool, contain, hold, approach, utility, usefulness
    
    \vspace{1em}
    Now, generate 10 sentences using a mix of both neutral and formal tones based on the provided input information.
    
    \end{tcolorbox}
    \caption{Example of prompt generated for tag \textit{T1006} for the \textit{CTI-to-MITRE} dataset.}
    \label{fig:data_aug_prompt} 
\end{figure}

\subsubsection{Fine-Tuning}
Following the data augmentation process described in the previous section, the classification component takes as input both the original and synthetic \gls{cti} sentences and assigns MITRE ATT\&CK technique labels to individual \gls{cti} sentences. This mapping is a crucial step in transforming unstructured threat intelligence into actionable, structured knowledge, helping \gls{cti} analysts to decide the best course of action to take.

We employ a transformer-based architecture adapted for multi-class classification, which takes as input a single \gls{cti} sentence and outputs a predicted technique label from the ATT\&CK framework. We chose transformer-based models due to their proven effectiveness in capturing contextual meaning. To adapt these models for classification, we modify their architecture by replacing the original language modeling head with a classification head suited to multi-class prediction over MITRE ATT\&CK techniques. This head typically includes a dropout layer followed by one or more linear layers, depending on the base architecture.  All model variants are sourced from the Hugging Face Transformers library, and detailed architectural specifications can be consulted in their official GitHub repository.\footnote{ \href{https://github.com/huggingface/transformers/blob/v4.52.3/src/transformers/models}{https://github.com/huggingface/transformers}} 

As already mentioned, the models are fine-tuned using Hugging Face’s Trainer API on a training set composed of both original \gls{cti} samples and synthetic data generated through our augmentation pipeline. This fine-tuning process enables the models to learn effective mappings between sentence-level threat descriptions and their corresponding ATT\&CK techniques. The synthetic data, in particular, addresses the long-tail distribution of ATT\&CK techniques by increasing the presence of low-frequency classes in the training set.

\subsection{System Deployment}
Once the \gls{llm} has been fine-tuned, the resulting \textit{MITRE Technique Classifier}, which internally includes this \gls{llm}, can be integrated into the workflow of \gls{cti} analysts. These fine-tuned models process new sentences extracted from \gls{cti} reports and automatically assign the corresponding MITRE ATT\&CK technique label.

In practice, this allows analysts to accelerate the technical analysis phase of threat reports. Rather than manually identifying and classifying each technique, analysts can rely on the classifier to provide the classification of the MITRE ATT\&CK techniques. Serving as a decision-support tool, the classifier proposes corresponding labels that analysts can validate and use to guide downstream actions.

Once the techniques have been identified, this information helps analysts prioritize threats, define targeted mitigation actions, and update internal databases or \gls{cti}-sharing platforms. In this way, the system not only accelerates the analysis process but also contributes directly to more informed and timely decision-making in real-world security operations.

To validate SynthCTI, we use real world \gls{cti} sentences datasets. Generating synthetic data to mitigate the impact of class imbalance, demonstrating significant enhancements in mapping threat descriptions to MITRE ATT\&CK techniques.

\section{Evaluation}\label{sec:evaluation}

We evaluate \textit{SynthCTI} through comprehensive experiments involving multiple models and settings. Given the class imbalance present in both \gls{cti} datasets (CTI-to-MITRE \cite{ctidataset} and \gls{tram} \cite{tramdataset} described in Section \ref{datasets}), even after applying augmentation. We adopt F1-macro as our primary evaluation metric. Unlike accuracy, which may be skewed by dominant classes, F1-macro gives equal weight to all classes by computing the F1-macro per class and averaging across them. This ensures that performance on underrepresented techniques is properly accounted for. For completeness, we also report accuracy, which offers insight into the overall correctness of predictions, even if it may mask performance on minority classes.

Table \ref{tab:models} summarises the selected models for our experiments. We have focused our efforts in small models that can be deployed locally with limited resources as, in a real world scenario, organizations may not be willing to share threat report data with third party services that offer the online large models or, to invest in the necessary infrastructure to run large on-premise models. 

\begin{table*}[h]
\centering
\caption{Models used.}
\label{tab:models}
\begin{tabular}{lccc}
\hline
 \textbf{Name}   & \textbf{Size} & \textbf{Fine Tuned} & \textbf{Available on}   \\
 \hline
albert-base-v2 \cite{Albert}  & 11,8M&     \ding{55}   &  https://huggingface.co/albert/albert-base-v2 \\

\hline

distilbert-base-uncased \cite{Sanh2019DistilBERTAD}   & 67M &    \ding{55}   &  https://huggingface.co/distilbert/distilbert-base-uncased \\

\hline
 bert-base-uncased \cite{Bert} & 110M &        \ding{55}& https://huggingface.co/google-bert/bert-base-uncased \\

\hline
 SecureBERT \cite{aghaei2023securebert}   & 125M &       \ding{51}   & https://huggingface.co/ehsanaghaei/SecureBERT \\

\hline
\end{tabular}
\end{table*}

A 40GB NVIDIA A100 GPU was used to run all experiments. For both datasets and all models, the hyperparameters used were a learning rate of $3e-05$, 10 epochs, a batch size of 32, a linear learning rate scheduler and an AdamW optimizer. The corresponding probabilities for the dropout layer in the classification head are the defaults, 0.1 for BERT, ABERT and SecureBERT (RoBERTa based model) and 0.2 for DistilBERT.

This section proceeds by detailing the specific datasets employed for our experiments. We then present a comprehensive comparison of various data augmentation methods, including \textit{SynthCTI}, followed by an analysis of the classification performance achieved across different models. Finally, we provide a deeper, qualitative and quantitative analysis of the data generated through our augmentation technique.

\subsection{Datasets}\label{datasets}
\textit{SynthCTI} makes use of two publicly available datasets that are directly aligned with the goal of classifying \gls{cti} sentences into MITRE ATT\&CK techniques: the \gls{tram} dataset \cite{tramdataset} and the CTI-to-MITRE dataset \cite{ctidataset}.
The \gls{tram} dataset was developed by the \gls{ctid}. \gls{tram} is an open-source platform designed to advance research in automating the mapping of \gls{cti} reports to the MITRE ATT\&CK framework. The dataset consists of sentences extracted from cyber threat intelligence reports that have been manually annotated. Like many real-world \gls{cti} sources, the \gls{tram} dataset exhibits class imbalance, with certain attack techniques being represented more frequently than others. Its sentence-level granularity makes it well-suited for training models that aim to assign technique labels at the sentence level.

The CTI-to-MITRE dataset was introduced in \cite{orbinato2022automatic}. This dataset contains natural language descriptions of \gls{cti} events, each labeled with the corresponding adversarial techniques from the MITRE ATT\&CK framework. It is distributed in CSV format and includes pairs of \gls{cti} sentences and their corresponding ATT\&CK technique labels. Similar to the \gls{tram} dataset, the CTI-to-MITRE dataset also suffers from class imbalance, reflecting the uneven frequency of techniques in operational threat reports. This dataset provides another valuable resource for training and evaluating models aimed at automating the mapping of \gls{cti} to the MITRE ATT\&CK framework. 

Both datasets are divided into two splits, one for training (80\%), and one for testing (20\%). The splits are stratified, to ensure that every class is represented in both training and test sets. It should be noted that the synthetic data produced using our augmentation method (described in Section \ref{dataug}) is added exclusively to the training split.

\subsection{Data augmentation method comparison}
\label{sec:augmentationcomparison}
To ensure a fair comparison across augmentation methods, we first determine the number of synthetic instances to generate per class. This is done by computing the average number of instances per class:

\[
\mu = \frac{1}{m}\sum_{i=1}^{m} N_i,
\]
where \(N_i\) is the number of examples in class \(i\)m and \(m\) is the total number of classes. 

To determine how many new instances to generate in each class with fewer examples than the average, we simply calculate the difference between \(\mu\) and the current class size, and set the difference to zero when this difference is negative:
\[
G_i = \max\bigl(0,\;\mu - N_i\bigr),
\]
where \(G_i\) is the number of synthetic instances to generate for class \(i\). 

\begin{figure}[h]
        \centering
        \includegraphics[width=1\linewidth]{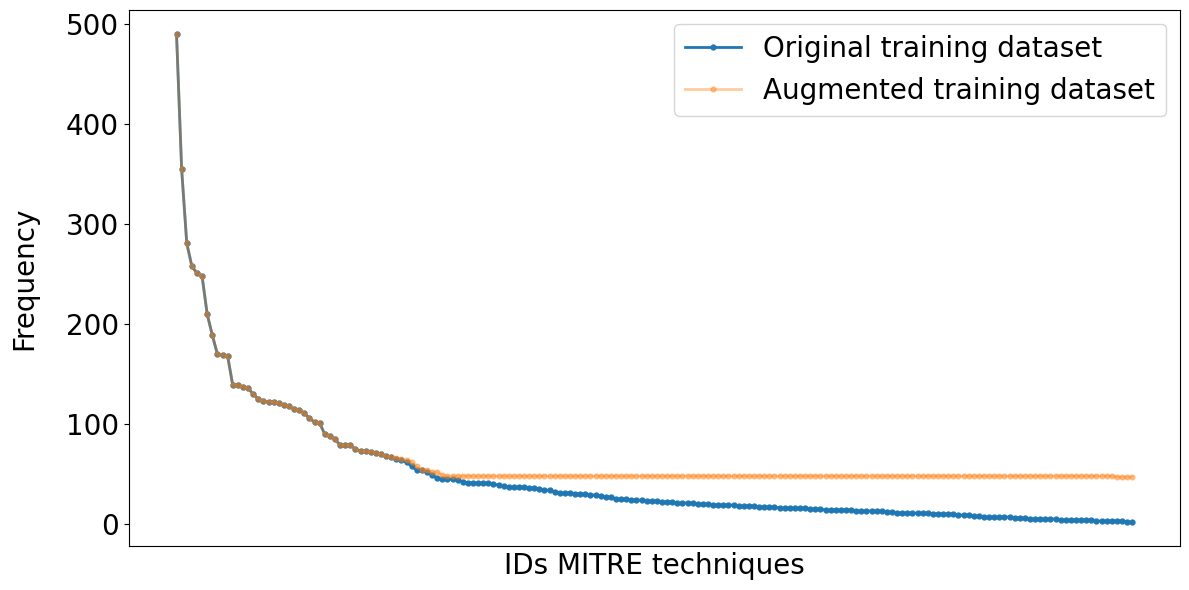} 
        \caption{Distribution of MITRE IDs from the CTI-to-MITRE dataset after joining the synthetic data.} 
        \label{fig:labels_aug_cti} 
\end{figure}

Figure~\ref{fig:labels_aug_cti} illustrates how this process balances the CTI-to-MITRE dataset by increasing underrepresented classes to the dataset average. Although not shown, the same balancing strategy provides similar results in the \gls{tram} dataset.

We selected baseline augmentation methods based on their simplicity, reproducibility, and low computational cost. Each method introduces different types of variations:

\begin{itemize}
    \item \textbf{Synonym replacement} \cite{tecnicasaug}: it replaces words with WordNet synonyms, introducing lexical variety while preserving semantics.
    \item \textbf{Random Swap} \cite{tecnicasaug}: it swaps the position of two words, causing minor syntactic change with little semantic impact.
    \item \textbf{MixUp for text} \cite{mixup}: it interpolates between two sentence embeddings to create hybrid synthetic samples.
    \item \textbf{Back-translation} \cite{backtranslation}: translation of the text into an intermediate language and back to the original, generating automatic paraphrases that maintain key semantics but introduce variations in structure and vocabulary.
    \item \textbf{Character noise} \cite{noise}: it introduces intentional spelling perturbations (e.g., typos) to simulate human error.
\end{itemize}

\begin{figure*}[h]
    \centering
    \includegraphics[width=1\linewidth]{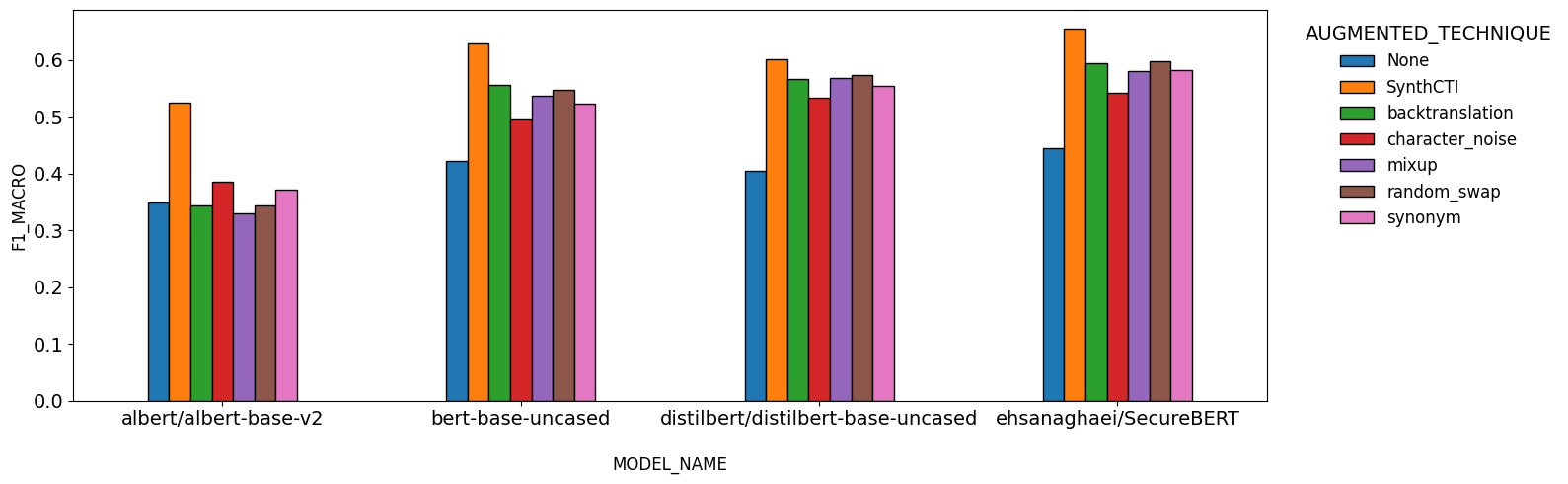} 
    \caption{Comparison of the F1-macro obtained for the different data augmentation techniques for the CTI-to-MITRE dataset.} 
    \label{fig:results_cti} 
\end{figure*}

\begin{figure*}[h]
    \centering
    \includegraphics[width=1\linewidth]{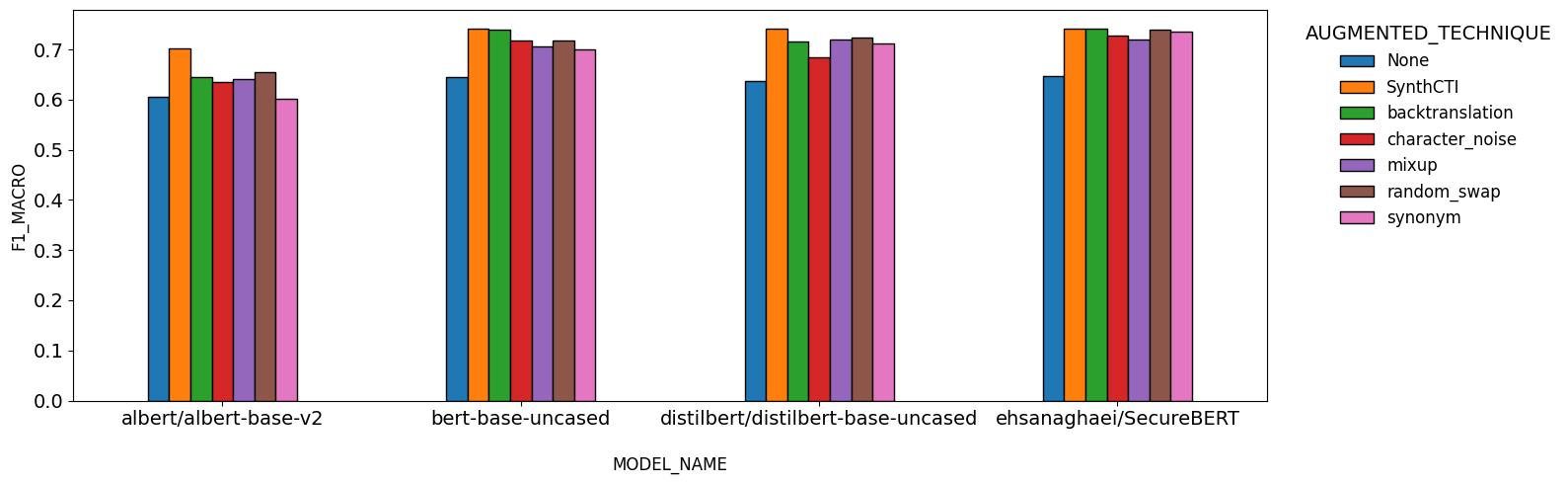} 
    \caption{Comparison of the F1-macro obtained for the different data augmentation techniques for the TRAM dataset.} 
    \label{fig:results_tram}
\end{figure*}

Figures~\ref{fig:results_cti} and~\ref{fig:results_tram} compare the F1-macro performance of all methods on both datasets.

For the CTI-to-MITRE dataset, SynthCTI shows the highest F1-macro scores across all models. The largest gain is with ALBERT (~0.52 vs. ~0.35 for others). BERT, SecureBERT, and DistilBERT also see improvements of 10+ points compared to their baseline augmentation scores. The ‘None’ baseline consistently underperforms, although ALBERT without augmentation still improves some augmentation techniques, highlighting its instability with noisy data.

In \gls{tram}, our method again achieves top performance, especially with ALBERT (0.70 vs. 0.62).  Larger models like BERT and SecureBERT see more modest but consistent improvements, often close to those of back-translation.

For both datasets, evaluation results show that, in low-capacity models like ALBERT and DistilBERT, SynthCTI provides substantial improvements, outperforming larger models trained without augmentation. This demonstrates that augmentation quality can compensate for model size. For instance, in the CTI-to-MITRE dataset, ALBERT with our augmentation method achieves an F1-macro of approximately 0.52, outperforming BERT and SecureBERT without augmentation, which reach only 0.42 and 0.44 respectively. Even in higher-capacity architectures, our method consistently improves results, confirming that augmentation benefits are not limited to small models. These results highlight the importance of semantic coherence and diversity in the generated data for training robust classifiers.

\subsection{Classification performance}
\begin{figure*}[htbp]
    \centering

    \begin{subfigure}[b]{0.45\textwidth}
        \includegraphics[width=\textwidth]{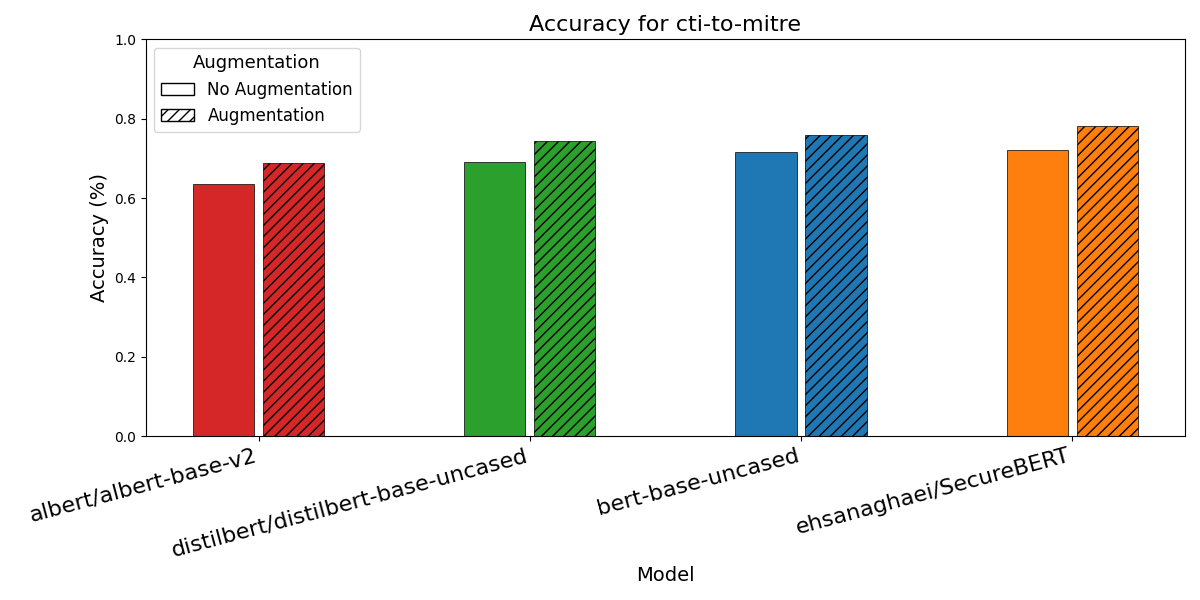}
        \caption{Accuracy results for CTI-to-MITRE.}
        \label{fig:accuracycti}
    \end{subfigure}
    \hfill
    \begin{subfigure}[b]{0.45\textwidth}
        \includegraphics[width=\textwidth]{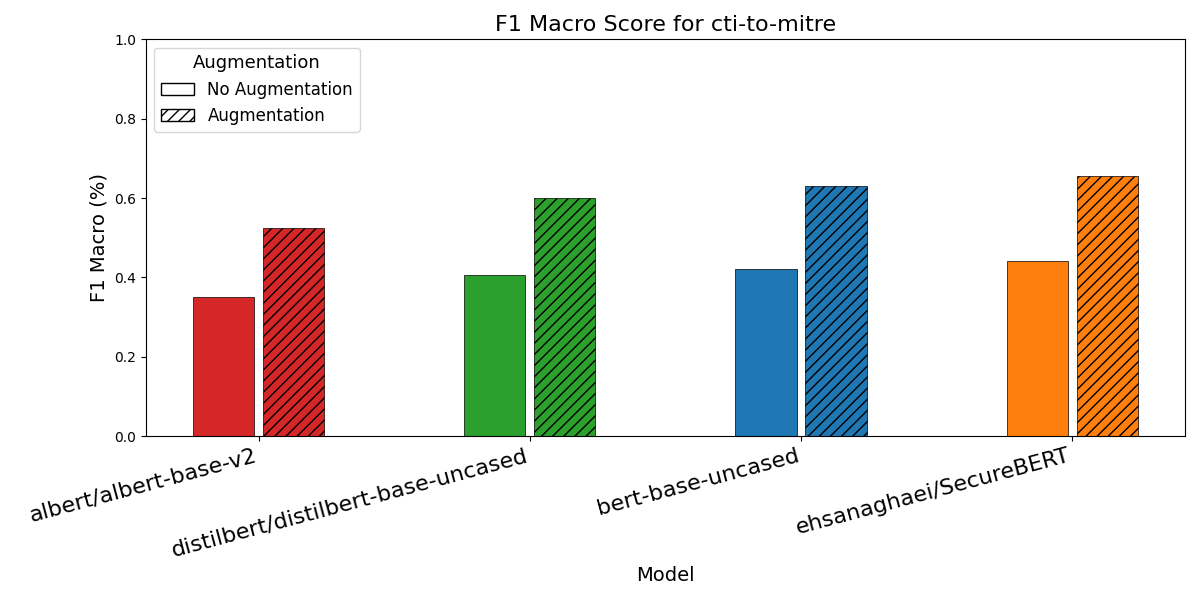}
        \caption{F1-macro results fo CTI-to-MITRE.}
        \label{fig:f1cti}
    \end{subfigure}

    \caption{Results for CTI-to-MITRE dataset.}
    \label{fig:resultsCTI}
\end{figure*}

\begin{figure*}[h]
    \centering

    \begin{subfigure}[b]{0.45\textwidth}
        \includegraphics[width=\textwidth]{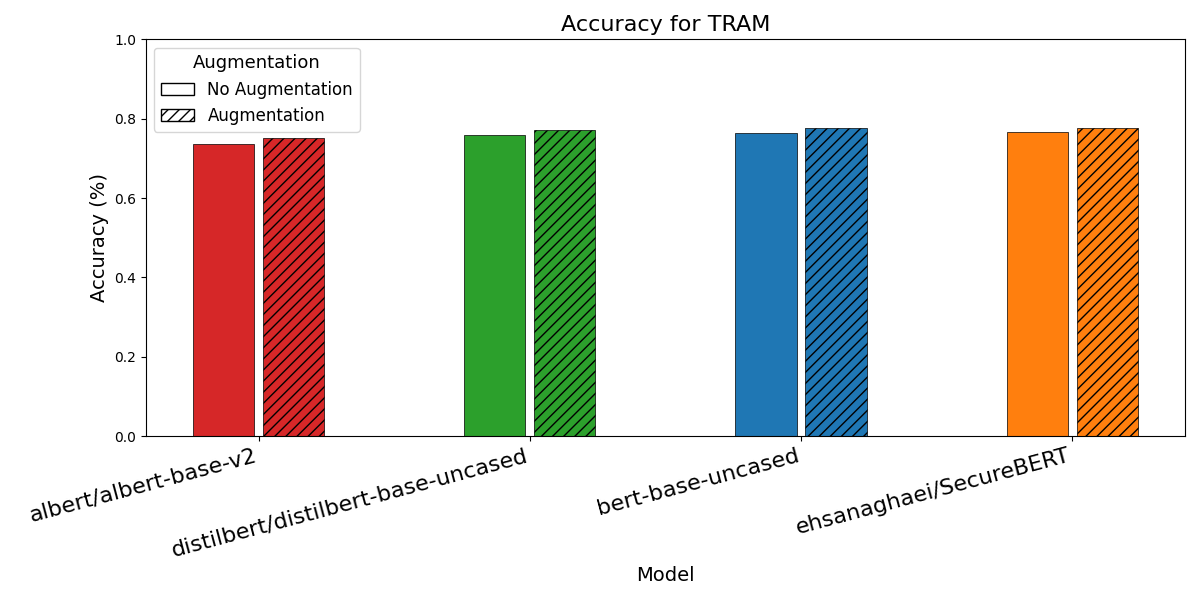}
        \caption{Accuracy results for \gls{tram}.}
        \label{fig:accuracyTRAM}
    \end{subfigure}
    \hfill
    \begin{subfigure}[b]{0.45\textwidth}
        \includegraphics[width=\textwidth]{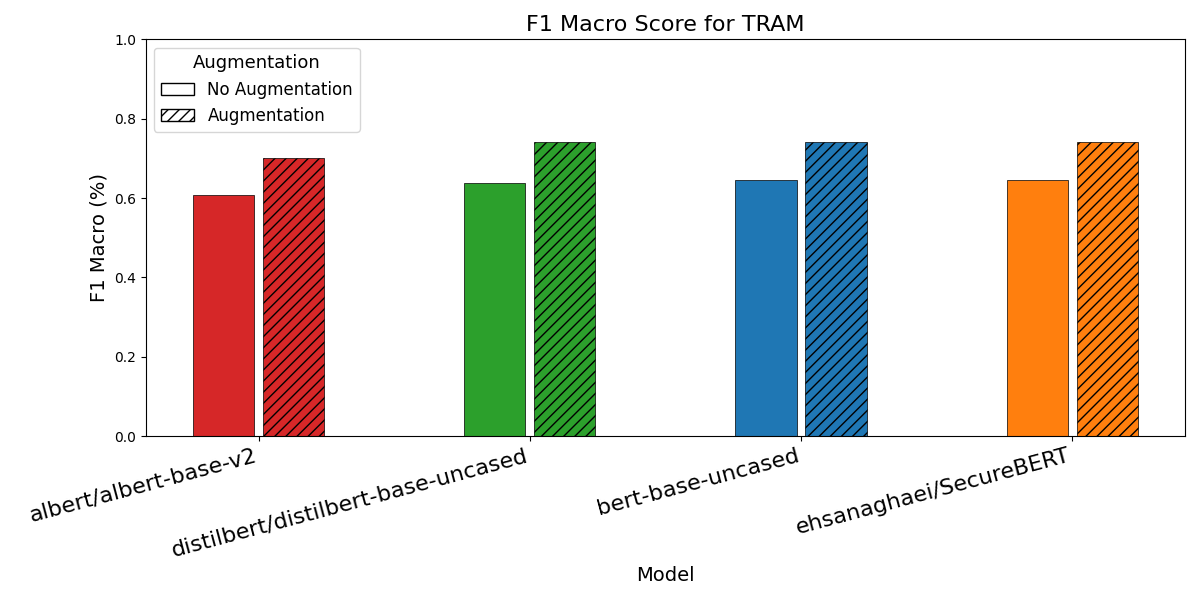}
        \caption{F1-macro results for \gls{tram}.}
        \label{fig:f1TRAM}
    \end{subfigure}
    \caption{Results for \gls{tram} dataset.}
    \label{fig:resultsTRAM}
\end{figure*}

Figures \ref{fig:resultsCTI} and \ref{fig:resultsTRAM} show the classification performance for CTI-to-MITRE and \gls{tram} datasets with and without applying our augmentation technique. As shown in both figures, the inclusion of augmented data consistently improves performance across all cases, particularly in terms of F1-macro score.

In the CTI-to-MITRE dataset (Figure \ref{fig:resultsCTI}), all models experience a notable increase in F1-macro when trained with augmented data. SecureBERT shows the best overall performance, reaching an F1-macro of 0.6558 with augmentation, compared to 0.4412 without. The second best performing model is BERT, with an F1-macro of 0.6302 when trained on augmented data. DistilBERT and ALBERT also follow this trend, improving their F1-macro scores by 48.36\% (from 0.4053 to 0.6013 ) and 50.36\% (from 0.3496 to 0.5256), respectively. Regarding accuracy, the improvements are less pronounced, but still consistent. SecureBERT improves from 0.7220 to 0.7811, while BERT reaches 0.7598. DistilBERT and ALBERT also experience around a 7.5\% increase in accuracy improving from 0.6915 to 0.7429 and 0.6363 to 0.688, respectively. These accuracy gains confirm that data augmentation leads to more robust classifiers across models. Furthermore, F1-macro improvements suggest that augmentation helps mitigate class imbalance, enhancing detection of minority classes, which represents a common challenge in \gls{cti} classification.

For the \gls{tram} dataset (Figure \ref{fig:resultsTRAM}), all models again benefit from augmentation, though with comparatively smaller gains. F1-macro increases are approximately  15\%, indicating a consistent but less dramatic impact. SecureBERT and DistilBERT are the best performing models at 0.7419 and 0.742 with augmentation, compared to 0.6468 and 0.6375 without, respectively. DistilBERT shows the largest gain on this dataset with a 16.40\% of improvement. This result underscores the value of augmentation for smaller models, helping them narrow the gap with larger ones. Accuracy gains on \gls{tram} were more modest (1–2 percentage points), showing that augmentation mainly improves detection of rare classes.

Comparing with previous work is challenging due to inconsistent metrics. Indeed, most report F1-weighted scores, limiting comparability \cite{orbinato2022automatic}, \cite{zhang-etal-2025-evaluating}. However, \cite{zhang-etal-2025-evaluating} reports F1-macro on the \gls{tram} dataset, enabling a direct comparison. Their 20-shot prompt-based method using llama-3-1-70b-instruct achieves an F1-macro of 0.708. Our method outperforms this with three out of four models: SecureBERT (0.7418), BERT (0.7412), and DistilBERT (0.7420). Even ALBERT, a lightweight 11M parameter model, matches their performance closely with 0.7013.

Our results also show a clear trend: with our approach, larger models generally achieve better results, but small models trained with augmented data can outperform bigger ones trained without it. For example, Albert (the smallest model among those evaluated) achieved a performance that, with the use of augmented data, exceeds models 10 times its size. This highlights that high-quality data can offset limited model capacity, offering a practical option in low-resource settings.

\begin{figure*}[h]
    \centering
    \includegraphics[width=1\linewidth]{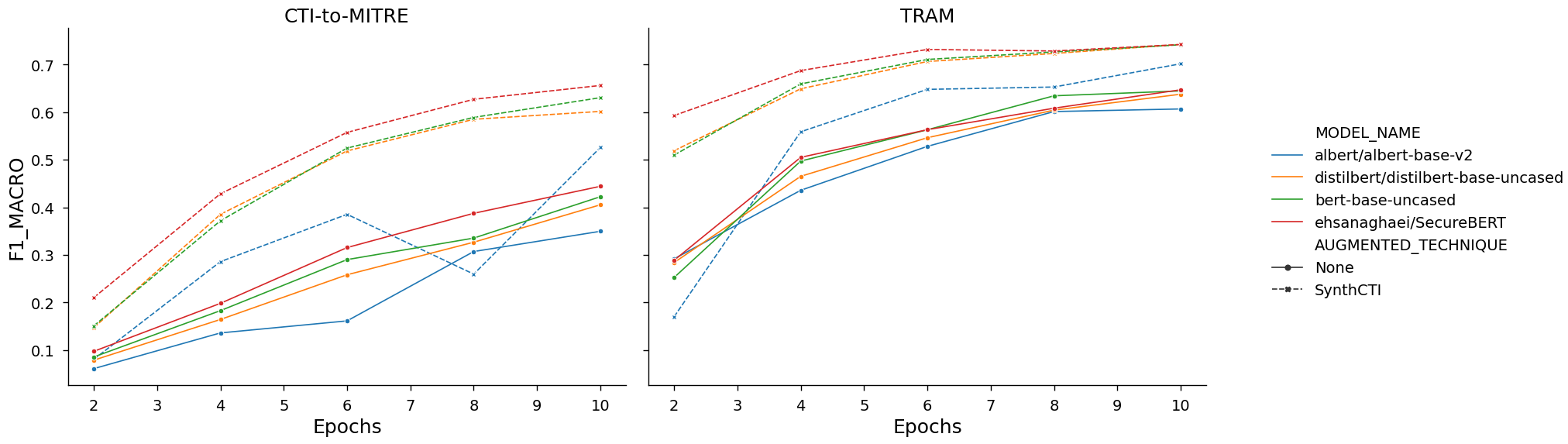} 
    \caption{Comparison of accuracy and F1-macro obtained for the different models for the CTI-to-MITRE and \gls{tram} datasets.} 
    \label{fig:comparacion_modelos}
\end{figure*}

Data augmentation also offers advantages early in training, as shown in Figure  \ref{fig:comparacion_modelos}. On the CTI-to-MITRE dataset, all models trained with augmented data learn faster and reach higher performance sooner. SecureBERT, for example, reaches an F1-macro of approximately 0.60 within just 6 epochs, which the non-augmented version does not reach even after 10. Similarly, for the \gls{tram} dataset, DistilBERT with augmentation achieves an F1-macro score above 0.60 by epoch 4, while the non-augmented DistilBERT only surpasses that value by epoch 10. This pattern holds across all models and both datasets. These results show that, in the \gls{cti} domain, synthetic data enables faster convergence, reducing training time while improving performance. This is especially relevant for security teams that need to retrain models regularly to keep up with new threats but operate under time or resource constraints. Indeed, more efficient fine-tuning cycles allow organizations to maintain updated classifiers without incurring high computational cost.

\subsection{Data augmentation analysis}
 We complement the quantitative results with a deeper analysis of the quality, structure, and diversity of the generated data.  This includes analyzing the distribution of synthetic embeddings, measuring semantic consistency, and analyzing potential limitations in classes with scarce data. Such analysis is provided in the subsections below.

 In addition to standard classification metrics, our analysis also involves clustering-based metrics to evaluate the structure and semantic quality of the generated data. Specifically, we use the Silhouette coefficient to measure intra-cluster cohesion and separation from other classes and the Davies–Bouldin (DB) index to evaluate the ratio between cluster compactness and separation. Cosine distance between original and synthetic embeddings is also used to quantify semantic similarity. These metrics provide a foundation for assessing augmentation quality in the next section.
 
\subsubsection{Latent Space Evaluation and Analysis}

We apply UMAP\cite{umap} to visualize how synthetic \gls{cti} sentences generated by our method are positioned in the embedding space relative to the original samples, allowing both global and local inspection of technique-level distributions. UMAP was chosen over t-SNE for its ability to preserve both local and global structure while being computationally more efficient. This aspect is specially relevant for large-scale \gls{cti} datasets. 

To assess the distribution quality of the synthetic samples, predefined thresholds are used to classify each class as either "strong" or "weak". This classification facilitates visual identification of clustering behavior in UMAP projections, such as cohesive halos, moderate dispersion, or chaotic scattering. Additionally, the quality of the generated data is assessed through a diversity–similarity analysis, comparing cosine distance and Self-BLEU scores to evaluate the trade-off between lexical novelty and semantic fidelity across classes.

\paragraph{Augmentation quality}
A low Silhouette score indicates poorly defined or overlapping clusters; similarly, a low Davies–Bouldin index implies compact and well-separated groups, while low cosine distance signals redundancy in generated content. To categorize the quality of synthetic data across classes, we define "strong" classes as those with a Silhouette $\geq$ 0.17, a DB index $<$ 2.0, and a cosine distance $\geq$ 0.10. Conversely, we considered "weak" classes with a Silhouette $<$ 0.05, a DB index $\geq$ 7.0, and a cosine distance $\le$ 0.03. These thresholds were empirically selected based on the observed distributions of these metrics across all classes in the datasets.

\begin{figure*}[h]
   \centering
   \begin{subfigure}[b]{0.8\linewidth}
       \centering
       \includegraphics[width=\linewidth]{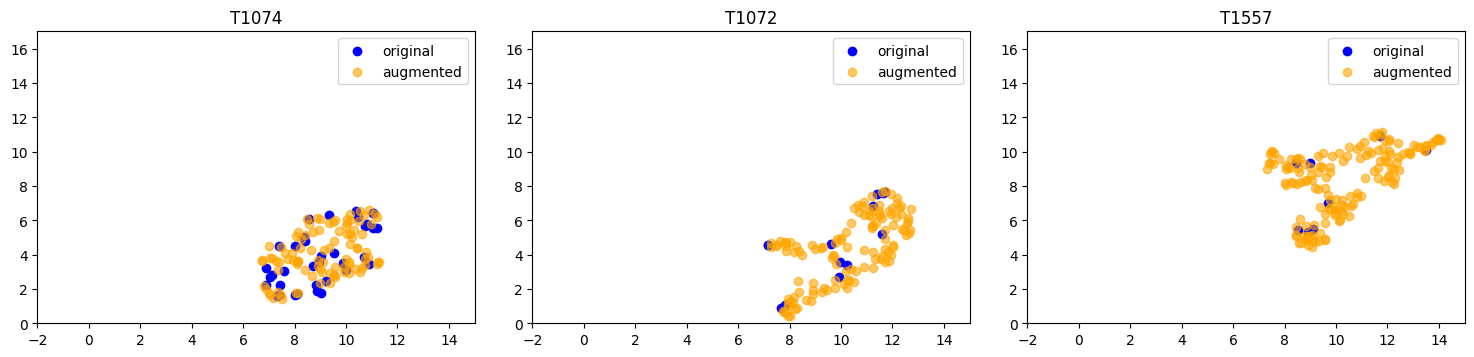}
       \caption{UMAP projection of embeddings for classes categorized as “weak” according to the defined thresholds.}
       \label{fig:debiles}
   \end{subfigure}
   \hfill
   \begin{subfigure}[b]{0.8\linewidth}
       \centering
       \includegraphics[width=\linewidth]{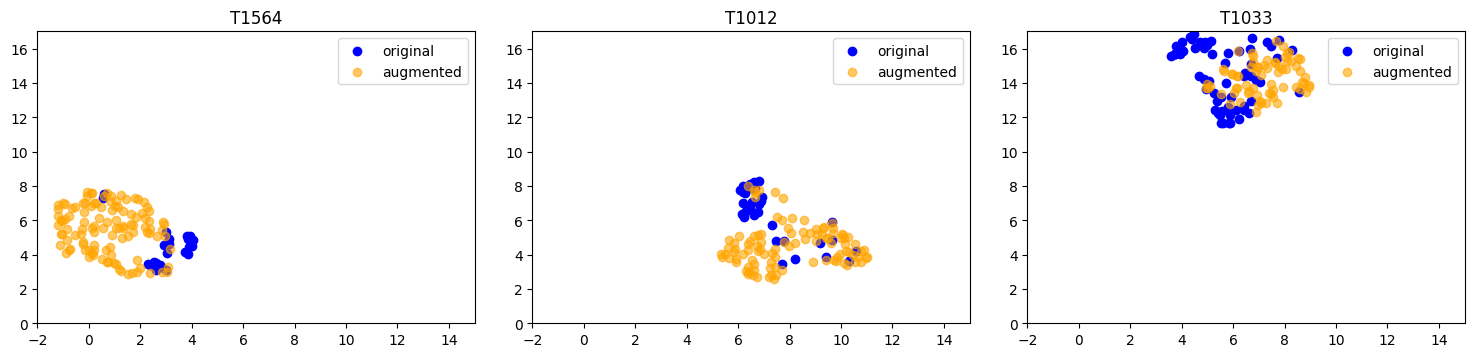}
       \caption{UMAP projection of embeddings for classes categorized as "strong" according to the defined thresholds.}
       \label{fig:fuertes}
   \end{subfigure}
   \caption{Analysis of synthetic data quality and embedding structure.}
   \label{fig:combined_analysis}
\end{figure*}

In the "strong" classes, such as T1564 (Figure ~\ref{fig:fuertes}), the synthetic points (orange) tend to form a cohesive cluster in proximity to the original instances (blue), often surrounding them or occupying nearby regions in the embedding space. This suggests that our method generates semantically coherent variations that remain within the class boundary while introducing useful diversity. Similar patterns can be observed in other strong classes like T1012 and T1033, although the degree of overlap may vary. These cases typically correspond to classes with a moderate number of original instances (around 20), which facilitates the generation of consistent augmentations.

In the "weak" classes shown in Figure ~\ref{fig:debiles}, such as T1074 (Silhouette 0.03, DB 15.6 and cosine 0.014), the synthetic examples appear chaotically dispersed or form disconnected subgroups with limited alignment to the original examples. This behavior is typically observed in classes with very few original samples (fewer than 10), where the \gls{llm} struggles to define a coherent semantic region, resulting in noisy or inconsistent augmentations.

\paragraph{Diversity vs Similarity}

\begin{figure}[h]
    \centering
    \includegraphics[width=1\linewidth]{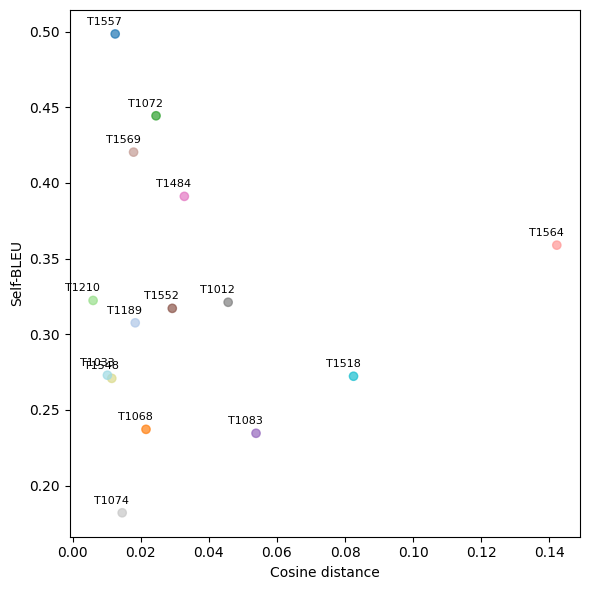} 
    \caption{Relationship between lexical novelty and semantic fidelity for the \gls{tram} set: cosine distance between the original and synthetic sentences versus Self-BLEU by class.} 
    \label{fig:diversity_vs_similarity}
\end{figure}

Figure~\ref{fig:diversity_vs_similarity} shows the relationship between diversity and similarity of the original sentences versus the synthetic ones, illustrating how the balance between semantic fidelity (orig–aug cosine distance) and lexical diversity (Self-BLEU) varies for each class. Some classes, such as T1012 (cosine distance 0.05, Self-BLEU 0.32), fall into the high-similarity/low-diversity region, meaning that while the generated samples retain the original semantics, they introduce limited lexical variety. Although semantically consistent, these augmentations may not provide sufficient variation to enhance model generalization.

Conversely, classes like T1557 exhibit high diversity (Self-BLEU $>$ 0.5) but very low semantic similarity (cosine 0.01), suggesting that while the sentences differ lexically, they may no longer reflect the original meaning, introducing semantic drift and potential noise. T1564, by contrast, displays well-balanced metrics (Self-BLEU 0.36, cosine distance 0.14), indicating that the generated sentences maintain semantic integrity while introducing meaningful lexical variety. This balance is desirable for improving model robustness without degrading label quality.

For underrepresented classes like T1074, low cosine distances and inconsistent Self-BLEU values suggest a lack of reliable augmentation patterns. In these cases, the generated sentences are either overly repetitive or exhibit uncontrolled variability, depending on the limited structure of the input data.

Overall, \textit{SynthCTI} produces synthetic data that, for most classes, achieves moderate cosine distances (around 0.4) and Self-BLEU values, balancing novelty and coherence. Nonetheless, in classes with extremely limited original examples, the quality and reliability of augmentations remain uncertain due to insufficient semantic grounding.

\subsubsection{Qualitative analysis}
We also perform a qualitative evaluation of some of the generated sentences. Our analysis focuses on those techniques that had very low representative examples in the initial datasets or those that produced relevant metrics during our quantitative analysis (Figures \ref{fig:fuertes} and \ref{fig:debiles}). To evaluate the quality of our synthetic data generation, we examined their semantic coherence, technical accuracy, and potential ambiguities in technique boundaries.

\paragraph{Dataset Size and Generation Quality} Our analysis reveals that generation quality is strongly correlated with original dataset size. Techniques with minimal representation (e.g., T1200 with 4 samples) exhibit overfitting to specific entities. For instance, the threat actor \enquote{DarkVishnya} appears in 30 out of 128 generated sentences (23\%), despite occurring in only one original example. Similarly, \enquote{raspberry} appears in 37 generated sentences, suggesting insufficient generalization from the original Raspberry Pi reference. This over-representation can result in models attributing specific threat actors directly to techniques rather than learning underlying tactical patterns. In T1557 (Adversary-in-the-Middle), ARP poisoning dominates over half of generated sentences, suggesting that our method may inadvertently prioritizes certain subtechniques over others, indicating a need for more balanced semantic theme selection.

In contrast, techniques with moderate representation show improved diversity. T1092 (Communication Through Removable Media) with 6 original examples demonstrates better balance, where specific tool names like \enquote{CHOPSTICK} appear in only 10 generated sentences (7\%) compared to 16\% in the original dataset. The increased variability and information in the original prompt dilutes very specific terms, indicating that training diversity effectively reduces over-representation.
For well-represented techniques such as T1012 (Registry Queries) with 85 examples, generated content maintains semantic coherence while expressing concepts through varied formulations.
For this particular technique, most generated sentences appropriately include registry-related terminology, facilitating classification while providing lexical diversity. However, specific malware names (Bankshot, Zeus, Carbon) appear in similar ratios to the original dataset, suggesting that even with larger datasets, entity-specific overfitting persists.

\paragraph{Technical Accuracy and Hallucination Issues} We identified instances where the \gls{llm} introduces technical inaccuracies when insufficient context is available. In T1072 (Software Deployment Tools), the model generates sentences that describe McAfee as \enquote{providing antivirus product protection} when the original sentences discuss a different product (a policy orchestrator) and references non-existent \enquote{administrative account logs}. McAfee appears in 23 out of 128 generated sentences despite being mentioned in only one original example, demonstrating how limited context can lead to technically incorrect content generation.

In T1189 (Drive-by Compromise), we observed information loss during generation. An original sentence describing \enquote{a Javascript based profiler called RICECURRY to profile a victim's web browser and deliver malicious code} was transformed to \enquote{the use of web-based profilers like ricecurry enables attackers to profile user browsing behavior}, removing the payload delivery component that defines the technique and confusing the profiling of a web browser with the profiling of the user behaviour within that web browser.

\paragraph{Technique Boundary Ambiguities} Previous research has shown how some of the boundaries between the techniques defined within the MITRE ATT\&CK can be ambiguous resulting in analyst assigning different techniques from the same description \cite{fayyazi2023uses}\cite{rege2023students}. The augmentation process occasionally reveals these inherent ambiguities between technique definitions. The sentences generated for T1092 (Communication Through Removable Media) exemplify this challenge. This technique requires adding removable media to target devices, potentially overlapping with T1200 (Hardware Additions). This can be seen as many of the mistakes made within T1092 are classifications into T1200.

\paragraph{Preprocessing and Generalization Strategies} Successful generalization strategies emerged from our pipeline's automatic preprocessing steps. The replacement of specific \gls{apt} numbers with generic \enquote{APT} references demonstrates effective entity normalization, as seen in T1092 where generated sentences reference \enquote{APTs strategy involved leveraging USB sticks} rather than specific groups like APT28. This approach could be extended using \gls{ner} to replace threat actor names, tool names, and other specific entities, potentially mitigating over-representation issues observed in smaller datasets. We leave this for future work.

\begin{figure}
    \centering
    \includegraphics[width=1\linewidth]{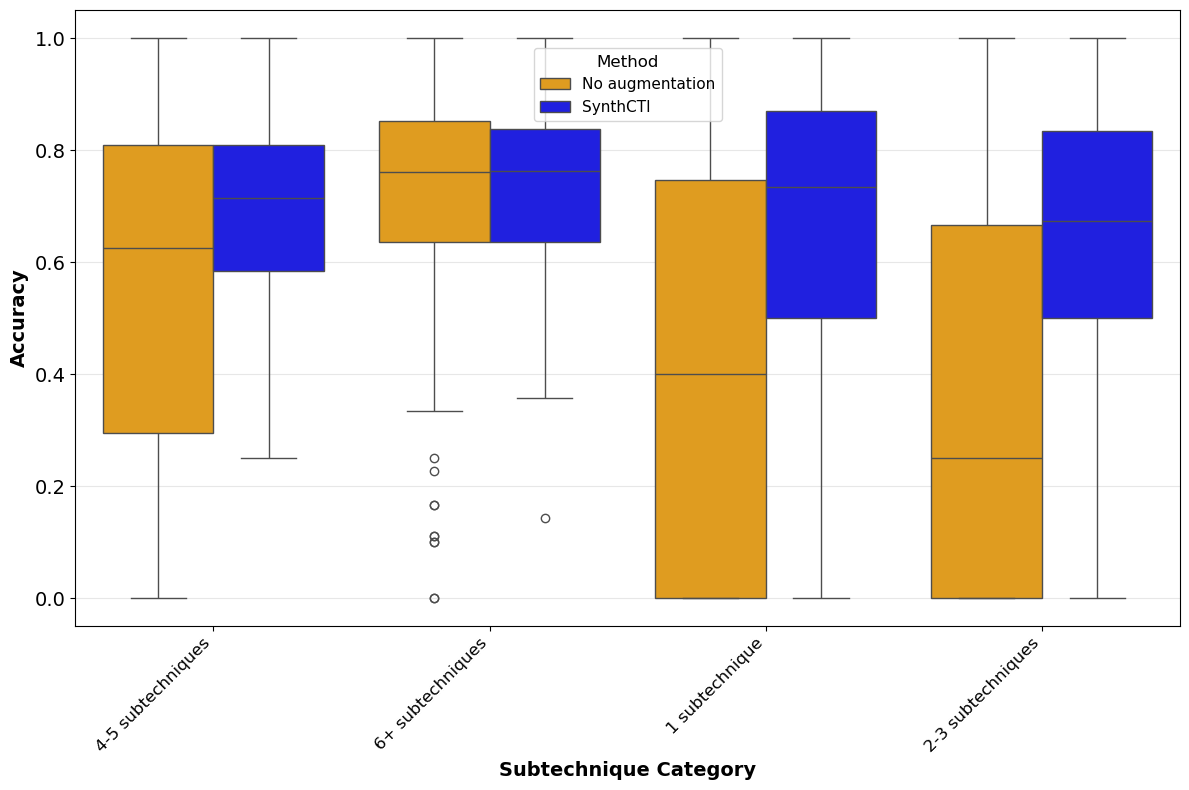}
    \caption{Classification performance per number of subtechniques.}
    \label{fig:subtechnique_acc_boxplot}
\end{figure}

\paragraph{Semantic Variation and Redundancy} Our examination reveals that many synthetic sentences are rephrasings of original examples with additional contextual details rather than genuinely novel expressions. In T1074 (Data Staging), generated content like \enquote{zebrocy stores all collected information in a single file before exfiltration streamlining the exfiltration process} closely mirrors the original \enquote{Zebrocy stores all collected information in a single file before exfiltration} with minimal semantic variation. While this maintains accuracy, it may limit the diversity needed for robust model training. We further analyse this issue by checking how the number of subtechniques for a particular technique can affect the performance of the resulting classifier. This is, we want to know how our augmentation affects the accuracy of the classifier depending on the amount of subtechniques we had to cover during our augmentation process.
The analysis of subtechnique performance (Figure~\ref{fig:subtechnique_acc_boxplot}) indicates that our augmentation method achieves affects more positively those techniques with fewer subtechniques. Single-subtechnique categories show visible improvements, while techniques with 4-5 subtechniques demonstrate more modest gains, suggesting our approach is particularly effective for well-defined, focused attack patterns.
The analysis confirms that while our \gls{llm}-based augmentation approach shows promise for enhancing \gls{cti} classification, the quality of synthetic generation is heavily dependent on original dataset richness. Techniques with fewer than 10 examples consistently exhibit problematic overfitting, while those with more than 20 examples demonstrate more balanced generation patterns, emphasizing the importance of sufficient training diversity for high-quality synthetic data generation.

\section{Conclusion} \label{sec:conclusion}
This work presents an \gls{llm}-based data augmentation pipeline that successfully improves on existing \gls{cti} sentence classification into MITRE ATT\&CK techniques. Our method consistently outperforms traditional augmentation techniques for all evaluated models. The approach shows relevant gains for smaller models, with ALBERT achieving F1-macro improvements from 0.35 to 0.52 on the CTI-to-MITRE dataset, suggesting that strategic data augmentation can reduce the need for extremely large models. We also showed that smaller models trained with augmented data converge to better performance more quickly, reducing computational costs and training time requirements. Moreover, our analysis shows that generation quality is correlated with the number of original samples per technique. Indeed, when fewer than 10 examples are available, generation tends to overfit specific entities, while well-represented techniques maintain both semantic coherence and lexical diversity. Our method proves especially effective for techniques with fewer subtechniques (i.e. well-defined, focused attack patterns). The augmentation process occasionally reveals inherent ambiguities between MITRE ATT\&CK technique definitions, highlighting opportunities for future work in refining label boundaries. Future work includes integrating entity normalization, and adapting the method to privacy-preserving scenarios through Federated Learning (FL), enabling collaborative CTI model training across organizations without exposing sensitive threat data.

\section*{Acknowledgments}
This study was supported by a Leonardo Grant 2023 to Researchers and Cultural Creators from the BBVA Foundation, the project CNS2023-145059 (INTEGRATOR) funded by MICIU/AEI/10.13039/501100011033 and the European Union NextGenerationEU/PRTR, the project PCI2023-145989-2 (REMINDER) funded by MI-CIU/AEI/10.13039/501100011033 and the European Union NextGenerationEU/PRTR, the MATM project (C127/23), with the collaboration of the Spanish Institue for Cybesecurity (INCIBE), and the Recovery, Transformation and Resilience Plan funded by the European Union (Next Generation), as well as the ONOFRE4 Project PID2023-148104OB-C43 funded by MICIU/AEI/10.13039/501100011033/ and FEDER/EU, and by the FPU Predoctoral Contract (FPU23/01816) granted by the Spanish Ministry of Science, Innovation and Universities.

\bibliographystyle{elsarticle-num} 

\bibliography{sample-base}

\begin{thebibliography}{10}
\expandafter\ifx\csname url\endcsname\relax
  \def\url#1{\texttt{#1}}\fi
\expandafter\ifx\csname urlprefix\endcsname\relax\def\urlprefix{URL }\fi
\expandafter\ifx\csname href\endcsname\relax
  \def\href#1#2{#2} \def\path#1{#1}\fi

\bibitem{alaeifar2024current}
P.~Alaeifar, S.~Pal, Z.~Jadidi, M.~Hussain, E.~Foo, Current approaches and future directions for cyber threat intelligence sharing: A survey, Journal of Information Security and Applications 83 (2024) 103786.

\bibitem{brown2023sans}
R.~Brown, K.~Nickels, Sans 2023 cti survey: Keeping up with a changing threat landscape, in: Tech. Rep, SANS Institute, 2023.

\bibitem{sun2023cyber}
N.~Sun, M.~Ding, J.~Jiang, W.~Xu, X.~Mo, Y.~Tai, J.~Zhang, Cyber threat intelligence mining for proactive cybersecurity defense: A survey and new perspectives, IEEE Communications Surveys \& Tutorials 25~(3) (2023) 1748--1774.

\bibitem{della2025cti}
S.~Della~Penna, R.~Natella, V.~Orbinato, L.~Parracino, L.~Pianese, Cti-hal: A human-annotated dataset for cyber threat intelligence analysis, arXiv preprint arXiv:2504.05866 (2025).

\bibitem{ctidataset}
{Dessert Lab}, {GitHub - dessertlab/cti-to-mitre-with-nlp}, \url{https://github.com/dessertlab/cti-to-mitre-with-nlp}.

\bibitem{tramdataset}
{Center for Threat-Informed Defense}, {center-for-threat-informed-defense/tram - GitHub}, \url{https://github.com/center-for-threat-informed-defense/tram}.

\bibitem{aghaei2023securebert}
E.~Aghaei, X.~Niu, W.~Shadid, E.~Al-Shaer, Securebert: A domain-specific language model for cybersecurity, in: Security and Privacy in Communication Networks: 18th EAI International Conference, SecureComm 2022, Virtual Event, October 2022, Proceedings, Springer, 2023, pp. 39--56.

\bibitem{orbinato2022automatic}
V.~Orbinato, M.~Barbaraci, R.~Natella, D.~Cotroneo, Automatic mapping of unstructured cyber threat intelligence: an experimental study:(practical experience report), in: 2022 IEEE 33rd International Symposium on Software Reliability Engineering (ISSRE), IEEE, 2022, pp. 181--192.

\bibitem{li2024automated}
L.~Li, C.~Huang, J.~Chen, Automated discovery and mapping att\&ck tactics and techniques for unstructured cyber threat intelligence, Computers \& Security 140 (2024) 103815.

\bibitem{you2024cyber}
W.~You, Y.~Park, Cyber-attack technique classification using two-stage trained large language models, arXiv preprint arXiv:2411.18755 (2024).

\bibitem{li2024empowering}
Y.~Li, K.~Ding, J.~Wang, K.~Lee, Empowering large language models for textual data augmentation, arXiv preprint arXiv:2404.17642 (2024).

\bibitem{dai2025auggpt}
H.~Dai, Z.~Liu, W.~Liao, X.~Huang, Y.~Cao, Z.~Wu, L.~Zhao, S.~Xu, F.~Zeng, W.~Liu, et~al., Auggpt: Leveraging chatgpt for text data augmentation, IEEE Transactions on Big Data (2025).

\bibitem{bayer2022multi}
M.~Bayer, T.~Frey, C.~Reuter, Multi-level fine-tuning, data augmentation, and few-shot learning for specialized cyber threat intelligence, Computers \& Security 134 (2023) 103430.

\bibitem{Campello2015}
R.~J. G.~B. Campello, D.~Moulavi, A.~Zimek, J.~Sander, \href{https://dl.acm.org/doi/10.1145/2733381}{Hierarchical density estimates for data clustering, visualization, and outlier detection}, ACM Transactions on Knowledge Discovery from Data 10~(1) (2015) 5:1--5:51.
\newline\urlprefix\url{https://dl.acm.org/doi/10.1145/2733381}

\bibitem{Albert}
Z.~Lan, M.~Chen, S.~Goodman, K.~Gimpel, P.~Sharma, R.~Soricut, \href{http://arxiv.org/abs/1909.11942}{{ALBERT:} {A} lite {BERT} for self-supervised learning of language representations}, CoRR abs/1909.11942 (2019).
\newblock \href {http://arxiv.org/abs/1909.11942} {\path{arXiv:1909.11942}}.
\newline\urlprefix\url{http://arxiv.org/abs/1909.11942}

\bibitem{ayoade2018automated}
G.~Ayoade, S.~Chandra, L.~Khan, K.~Hamlen, B.~Thuraisingham, Automated threat report classification over multi-source data, in: 2018 IEEE 4th International Conference on Collaboration and Internet Computing (CIC), IEEE, 2018, pp. 236--245.

\bibitem{legoy2020automated}
V.~Legoy, M.~Caselli, C.~Seifert, A.~Peter, Automated retrieval of att\&ck tactics and techniques for cyber threat reports, arXiv preprint arXiv:2004.14322 (2020).

\bibitem{zhang-etal-2025-evaluating}
B.~Zhang, M.~Takeuchi, R.~Kawahara, S.~Asthana, M.~M. Hossain, G.-J. Ren, K.~Soule, Y.~Mai, Y.~Zhu, \href{https://aclanthology.org/2025.naacl-industry.40/}{Evaluating large language models with enterprise benchmarks}, in: W.~Chen, Y.~Yang, M.~Kachuee, X.-Y. Fu (Eds.), Proceedings of the 2025 Conference of the Nations of the Americas Chapter of the Association for Computational Linguistics: Human Language Technologies (Volume 3: Industry Track), Association for Computational Linguistics, Albuquerque, New Mexico, 2025, pp. 485--505.
\newblock \href {https://doi.org/10.18653/v1/2025.naacl-industry.40} {\path{doi:10.18653/v1/2025.naacl-industry.40}}.
\newline\urlprefix\url{https://aclanthology.org/2025.naacl-industry.40/}

\bibitem{feng2021}
S.~Y. Feng, V.~Gangal, J.~Wei, S.~Chandar, S.~Vosoughi, T.~Mitamura, E.~Hovy, A survey of data augmentation approaches for nlp, in: Findings of ACL-IJCNLP, Association for Computational Linguistics, 2021, pp. 968--988.

\bibitem{cuong2025towards}
H.~Cuong~Nguyen, S.~Tariq, M.~Baruwal~Chhetri, B.~Quoc~Vo, Towards effective identification of attack techniques in cyber threat intelligence reports using large language models, in: Companion Proceedings of the ACM on Web Conference 2025, 2025, pp. 942--946.

\bibitem{beltagy2019scibert}
I.~Beltagy, K.~Lo, A.~Cohan, Scibert: A pretrained language model for scientific text, arXiv preprint arXiv:1903.10676 (2019).

\bibitem{gemma3}
A.~Kamath, J.~Ferret, S.~Pathak, N.~Vieillard, R.~Merhej, S.~Perrin, T.~Matejovicova, A.~Ram{\'e}, M.~Rivi{\`e}re, L.~Rouillard, et~al., \href{https://doi.org/10.48550/arXiv.2503.19786}{Gemma 3 technical report}, CoRR abs/2503.19786 (March 2025).
\newline\urlprefix\url{https://doi.org/10.48550/arXiv.2503.19786}

\bibitem{Wang2020}
W.~Wang, F.~Wei, L.~Dong, H.~Bao, N.~Yang, M.~Zhou, Minilm: deep self-attention distillation for task-agnostic compression of pre-trained transformers (2020).

\bibitem{Reimers2019}
N.~Reimers, I.~Gurevych, \href{https://doi.org/10.18653/v1/D19-1410}{Sentence-bert: Sentence embeddings using siamese bert-networks}, Proceedings of the 2019 Conference on Empirical Methods in Natural Language Processing (2019).
\newline\urlprefix\url{https://doi.org/10.18653/v1/D19-1410}

\bibitem{Ester1996}
M.~Ester, H.-P. Kriegel, J.~Sander, X.~Xu, \href{https://dl.acm.org/doi/10.5555/3001460.3001507}{A density-based algorithm for discovering clusters in large spatial databases with noise}, in: Proceedings of the 2nd International Conference on Knowledge Discovery and Data Mining, 1996, pp. 226--231.
\newline\urlprefix\url{https://dl.acm.org/doi/10.5555/3001460.3001507}

\bibitem{lda}
D.~M. Blei, A.~Y. Ng, M.~I. Jordan, \href{http://www.jmlr.org/papers/v3/blei03a.html}{Latent dirichlet allocation}, Journal of Machine Learning Research 3 (2003) 993--1022.
\newline\urlprefix\url{http://www.jmlr.org/papers/v3/blei03a.html}

\bibitem{keybert}
M.~Grootendorst, \href{https://doi.org/10.5281/zenodo.4461265}{Keybert: Minimal keyword extraction with bert.} (2020).
\newblock \href {https://doi.org/10.5281/zenodo.4461265} {\path{doi:10.5281/zenodo.4461265}}.
\newline\urlprefix\url{https://doi.org/10.5281/zenodo.4461265}

\bibitem{Miller1995}
G.~A. Miller, \href{https://dl.acm.org/doi/10.1145/219717.219748}{Wordnet: A lexical database for english}, Communications of the ACM 38~(11) (1995) 39--41.
\newblock \href {https://doi.org/10.1145/219717.219748} {\path{doi:10.1145/219717.219748}}.
\newline\urlprefix\url{https://dl.acm.org/doi/10.1145/219717.219748}

\bibitem{moreno2016large}
I.~Moreno-S{\'a}nchez, F.~Font-Clos, {\'A}.~Corral, Large-scale analysis of zipf’s law in english texts, PloS one 11~(1) (2016) e0147073.

\bibitem{Flesch}
R.~Flesch, A new readability yardstick, Journal of Applied Psychology 32~(3) (1948) 221--233.
\newblock \href {https://doi.org/10.1037/h0057532} {\path{doi:10.1037/h0057532}}.

\bibitem{Gunning}
R.~Gunning, The Technique of Clear Writing, revised Edition, McGraw-Hill, New York, 1952, introduces the Gunning Fog Index for measuring readability.

\bibitem{Sanh2019DistilBERTAD}
V.~Sanh, L.~Debut, J.~Chaumond, T.~Wolf, Distilbert, a distilled version of bert: smaller, faster, cheaper and lighter, ArXiv abs/1910.01108 (2019).

\bibitem{Bert}
J.~Devlin, M.~Chang, K.~Lee, K.~Toutanova, \href{http://arxiv.org/abs/1810.04805}{{BERT:} pre-training of deep bidirectional transformers for language understanding}, CoRR abs/1810.04805 (2018).
\newblock \href {http://arxiv.org/abs/1810.04805} {\path{arXiv:1810.04805}}.
\newline\urlprefix\url{http://arxiv.org/abs/1810.04805}

\bibitem{tecnicasaug}
J.~Wei, K.~Zou, \href{https://aclanthology.org/D19-1670/}{{EDA}: Easy data augmentation techniques for boosting performance on text classification tasks}, in: K.~Inui, J.~Jiang, V.~Ng, X.~Wan (Eds.), Proceedings of the 2019 Conference on Empirical Methods in Natural Language Processing and the 9th International Joint Conference on Natural Language Processing (EMNLP-IJCNLP), Association for Computational Linguistics, Hong Kong, China, 2019, pp. 6382--6388.
\newblock \href {https://doi.org/10.18653/v1/D19-1670} {\path{doi:10.18653/v1/D19-1670}}.
\newline\urlprefix\url{https://aclanthology.org/D19-1670/}

\bibitem{mixup}
H.~Zhang, M.~Cisse, Y.~N. Dauphin, D.~Lopez-Paz, \href{https://openreview.net/forum?id=r1Ddp1-Rb}{mixup: Beyond empirical risk minimization}, in: International Conference on Learning Representations (ICLR), 2018.
\newline\urlprefix\url{https://openreview.net/forum?id=r1Ddp1-Rb}

\bibitem{backtranslation}
R.~Sennrich, B.~Haddow, A.~Birch, \href{https://aclanthology.org/P16-1009/}{Improving neural machine translation models with monolingual data}, in: K.~Erk, N.~A. Smith (Eds.), Proceedings of the 54th Annual Meeting of the Association for Computational Linguistics (Volume 1: Long Papers), Association for Computational Linguistics, 2016, pp. 86--96.
\newblock \href {https://doi.org/10.18653/v1/P16-1009} {\path{doi:10.18653/v1/P16-1009}}.
\newline\urlprefix\url{https://aclanthology.org/P16-1009/}

\bibitem{noise}
Y.~Belinkov, Y.~Bisk, \href{https://openreview.net/forum?id=BJ8vJebC-}{Synthetic and natural noise both break neural machine translation}, in: International Conference on Learning Representations(ICLR), 2018.
\newline\urlprefix\url{https://openreview.net/forum?id=BJ8vJebC-}

\bibitem{umap}
L.~McInnes, J.~Healy, N.~Saul, L.~Großberger, \href{https://doi.org/10.21105/joss.00861}{Umap: Uniform manifold approximation and projection}, Journal of Open Source Software 3~(29) (2018) 861.
\newblock \href {https://doi.org/10.21105/joss.00861} {\path{doi:10.21105/joss.00861}}.
\newline\urlprefix\url{https://doi.org/10.21105/joss.00861}

\bibitem{fayyazi2023uses}
R.~Fayyazi, S.~J. Yang, On the uses of large language models to interpret ambiguous cyberattack descriptions, arXiv preprint arXiv:2306.14062 (2023).

\bibitem{rege2023students}
A.~Rege, J.~Williams, R.~Bleiman, K.~Williams, Students’ application of the {MITRE} {ATT\&CK}\textregistered\ framework via a real‑time cybersecurity exercise, in: Proceedings of the 22nd European Conference on Cyber Warfare and Security (ECCWS), Vol.~22, Academic Conferences International Limited, Piraeus, Greece, 2023, pp. 384--394.
\newblock \href {https://doi.org/10.34190/eccws.22.1.1126} {\path{doi:10.34190/eccws.22.1.1126}}.

\end{thebibliography}

\end{document}